\documentclass[11pt,a4paper]{article}

\usepackage{jheppub}
\usepackage{amssymb,amsmath}
\usepackage{graphicx}
\usepackage{slashed}
\usepackage[normalem]{ulem}
\usepackage[caption=false]{subfig}

\def \tr {\mathop{\rm tr}\nolimits}

\newcommand \ket [1] {|{#1}\rangle}
\newcommand \bra [1] {\langle {#1}|}

\newcommand{\pa}{\partial}

\newcommand{\be}{\begin{equation}}
\newcommand{\ee}{\end{equation}}
\newcommand{\bea}{\begin{eqnarray}}
\newcommand{\eaa}{\end{eqnarray}}

\newcommand{\nt}{\nonumber\\}

\renewcommand{\a}{\alpha}

\newcommand{\da}{{\dot\alpha}}
\newcommand{\db}{{\dot\beta}}
\newcommand{\tl}{{\tilde\lambda}}

\newcommand{\tx}{{\tilde x}}
\newcommand{\tb}{{\tilde b}}

\renewcommand{\b}{\beta}

\newcommand{\la}{\lambda}

\newcommand{\ep}{\epsilon}
\newcommand{\vep}{\varepsilon}

\newcommand{\cN}{{\cal N}}

\newcommand{\p}[1]{(\ref{#1})}
\newcommand{\bt}[1]{{\bar t}}

\newcommand \vev [1] {\langle{#1}\rangle}

\def\eps{\epsilon}

\allowdisplaybreaks[3]

\makeatletter
\renewcommand\@fpheader{} 
\renewcommand\@journal{}
\makeatother

\title{Implications of nonplanar dual conformal symmetry}

\preprint{CERN-TH-2018-168, LAPTH-029/18, MPP-2018-176, MITP/18-061}

\author{D.\ Chicherin$^a$, J.\ M.\ Henn$^{a,b}$, E.\ Sokatchev$^{a,c}$}

\affiliation{
$^a$ 
PRISMA Cluster of Excellence, Johannes Gutenberg University, 55099 Mainz, Germany\\
$^b$ Max-Planck-Institut f{\"u}r Physik, Werner-Heisenberg-Institut, 80805 M{\"u}nchen, Germany\\
$^c$ LAPTh, Universit\'e Savoie Mont Blanc, CNRS, B.P. 110, F-74941 Annecy-le-Vieux, France}

\emailAdd{chicherin@uni-mainz.de}
\emailAdd{henn@uni-mainz.de}
\emailAdd{emeri.sokatchev@cern.ch}

\abstract{
Recently, Bern et al observed that a certain class of next-to-planar Feynman integrals possess a bonus symmetry that is closely related to dual conformal symmetry. 
It corresponds to a projection of the latter along a certain lightlike direction.
Previous studies were performed at the level of the loop integrand, and a Ward identity for the integral was formulated. We investigate the implications of the symmetry at the level of the integrated quantities. In particular, we focus on the phenomenologically important case of five-particle scattering. The symmetry simplifies the four-variable problem to a three-variable one. In the context of the recently proposed  space of pentagon functions, the symmetry is much stronger. We find that it drastically reduces the allowed function space, leading to a well-known space of three-variable functions. Furthermore, we show how to use the symmetry in the presence of infrared divergences,  where one obtains an anomalous Ward identity.  We verify that the Ward identity is satisfied by the leading and subleading poles of several nontrivial five-particle integrals. 
Finally, we present examples of integrals that possess both ordinary and dual conformal symmetry.
 }

\begin{document}
\unitlength1cm
\maketitle

\newpage

\section{Introduction}
\label{sec:intro}

Scattering amplitudes are fascinating objects that are important in collider physics, and
at the same time are of theoretical interest, as their study allows to uncover novel features of quantum field theory.
The maximally supersymmetric Yang-Mills theory ($\mathcal{N}=4$ sYM) has emerged as a theoretical laboratory for this.
Many of the new ideas found there were later generalized and are being used for QCD calculations,
for example. Despite enormous progress in finding new features of scattering amplitudes, and novel ways of 
computing them, the bulk of the progress was made in the planar  sector of the theory. 

This is reflected by the fact that very few nonplanar amplitudes are known explicitly.
At the level of integrated quantities, only the four-particle amplitude is known at two and three loops \cite{Henn:2016jdu}.
At the level of the loop integrand, the two-loop five-particle amplitude is known \cite{Carrasco:2011mn,Bern:2015ple}.

An important feature of planar scattering amplitudes in $\mathcal{N}=4$ sYM 
is that they have a hidden dual (super)conformal and Yangian symmetry \cite{Drummond:2006rz,Alday:2007hr,Drummond:2007aua,Drummond:2007au,Drummond:2008vq,Berkovits:2008ic,Beisert:2008iq,Brandhuber:2008pf,Drummond:2009fd}. Its discovery was instrumental for many further developments in the theory. It is an open question whether this symmetry also manifests itself in some form at the level of nonplanar scattering amplitudes.

Recently, Bern et al \cite{Bern:2017gdk,Bern:2018oao} found that a certain class of nonplanar Feynman integrals have a bonus symmetry that is closely related to dual conformal symmetry. 
The class of integrals they consider can be called `next-to-planar': these are graphs that can be made planar upon removing one of the external legs. In this way one obtains an associated planar graph, for which dual conformal transformations can be defined unambiguously. The authors show that the original `next-to-planar' integral can still be invariant under a subset of dual conformal transformations, namely those projected along the direction of the lightlike momentum of the leg that was removed. 
We call this symmetry {\it directional dual conformal invariance} (DDCI).
The rules for constructing integrands that are covariant under these transformations are very similar to the planar case, with a few new features.
An important open question is how powerful this DDCI is.

The authors of \cite{Bern:2017gdk,Bern:2018oao} also formulated Ward identities for this DDCI. For integrals having infrared divergences (ultraviolet divergences are excluded by the dual conformal power counting), the Ward identities are anomalous. In order to make use of them, the anomalous term has to be evaluated. This is to be contrasted with the case of planar amplitudes, where the anomaly is known to all loop orders \cite{Drummond:2007au}.

In the present paper we address these open questions.
We investigate in detail the implications of the DDCI for the integrated quantities.

We discuss in general the construction of covariants and invariants of the DDCI, highlighting differences to the planar case. Then, we focus on the important case of five-particle integrals and amplitudes. The partial dual conformal symmetry eliminates one of the four dimensionless kinematic invariants, leaving three DDCI variables. We discuss different useful choices of theses variables. The reduction from four to three variable functions may not seem a very strong constraint. However, it becomes so when combined with the knowledge about the space of allowed integral functions. In Ref.~\cite{Chicherin:2017dob}
it was conjectured, based on the planar result of \cite{Gehrmann:2015bfy}, that nonplanar five-particle integrals evaluate to pentagon functions  characterized by a 31-letter alphabet.  Here we show that only a subset of 10 letters is compatible with the requirement of DDCI along the direction of, e.g, leg $p_3$. This drastically reduces the space of allowed functions. It turns out that the restricted 10-letter alphabet is well known from other studies \cite{Brown:2009qja}. In particular, it has appeared in the six-point fully dual conformal planar amplitude \cite{Goncharov:2010jf}.  

After this general investigation of the function space, we present a number of concrete examples of two-loop next-to-planar integrals. We want to demonstrate how they satisfy the (anomalous) DDCI Ward identity. The leading pole of a divergent integral is exactly DDCI. This is a nontrivial statement, since the leading poles of our integrals are given by sophisticated weight two and three hyperlogarithmic functions. We evaluate them and show that their symbols are expressed in terms of the 10-letter alphabet. 

Furthermore, we analyze the structure of the anomalous DDCI Ward identity at the next, subleading level. To this end,  we evaluate the leading term of the anomaly and compare it to the directional conformal transformation of the subleading term in the integral. We show perfect agreement of the symbols, now given by the full 31-letter alphabet. 

The paper is organized as follows.
In Section \ref{sec:bonus} we review the definition of dual conformal transformations for next-to-planar Feynman graphs,
their most important properties, and the anomalous Ward identities.
In Section \ref{sec:implications-pentagon-functions} we analyze in detail the implications for the integrated functions,
using five-particle scattering amplitudes as our main example.
We conclude and discuss the results in Section \ref{sec:discussion}.
There are two appendices. Appendix \ref{aA} reviews the construction of conformal covariants and invariants, and then discusses the new features inherent to the subset of dual conformal transformations used in the nonplanar case. 
Appendix \ref{sec:six-dim-examples} contains examples of finite six-dimensional DDCI integrals.

\section{Directional dual conformal invariance of next-to-planar Feynman integrals}
\label{sec:bonus}

\begin{figure}
\begin{center}
\includegraphics[width=7cm]{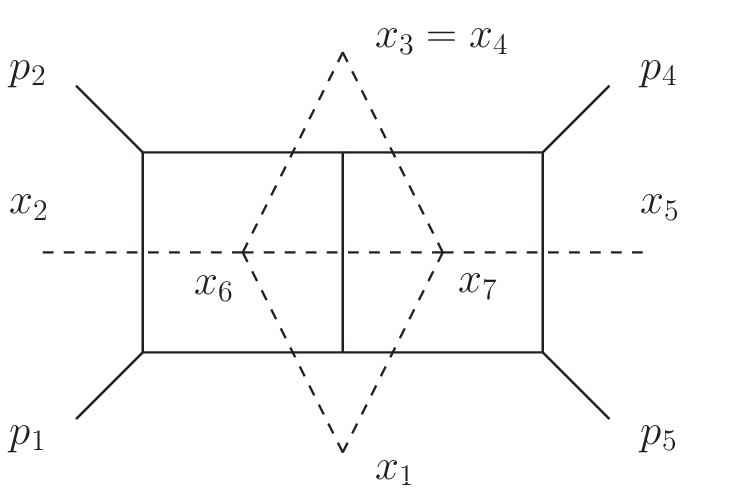} \quad \includegraphics[width=7cm]{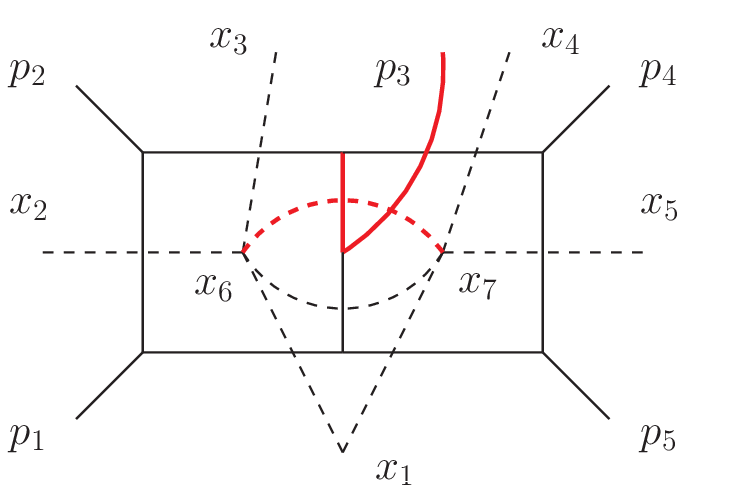}
\end{center}
\caption{Dual space for a planar graph and its nonplanar analog obtained by adding one leg.}
\label{fig:dualcoords}
\end{figure}

The notion of dual conformal invariance (DCI) for a Feynman integral relies on its dual space description. For an integral depending on $n$ external momenta $p_i$ (not necessarily lightlike), $I(p_1,\ldots,p_n)$, the  dual coordinates can be defined, e.g., as follows:
\begin{align}\label{1.1}
& p_i = x_{i+1, i}\equiv x_{i+1}-x_i \quad {\rm with} \ x_{n+1} \equiv x_1  \qquad  \Rightarrow \sum_{i=1}^n p_i =0   \,.
\end{align}
This is simply a way of solving the momentum conservation condition for the external momenta, it can be used for planar as well as nonplanar integrals. The difference between the two topologies appears at the level of the internal lines (propagators) involving the loop momenta.  This is illustrated in Fig.~\ref{fig:dualcoords}. The diagram on the left is planar, that on the right is nonplanar. The former can be obtained from the latter by removing the leg $p_3$.\footnote{We keep the convention of Ref.~\cite{Bern:2018oao} to use leg $p_3$ for creating nonplanar topologies, but we prefer a different labeling of the dual points, see \p{1.1}.} Equivalently, in terms of the dual coordinates we identify points $x_4=x_3$ and obtain a dual space picture in which each dashed line connecting two dual points crosses one and only one propagator line. This guarantees that all the  propagators can be put in the dual conformal form $1/x_{ij}^2$. So, the expression for the left diagram is
\begin{align}\label{2.2}
I_{\rm pl} (p_1,p_2,p_4,p_5) = I_{\rm pl} (x_1,x_2,x_3=x_4,x_5) = \int \frac{d^D x_6 d^Dx_7\, N_{\rm pl}(p_i)}{x^2_{16}x^2_{26}x^2_{36} x^2_{67} x^2_{37}x^2_{57}x^2_{17}}\,.
\end{align}
Here we use a dimensionally regularized measure with $D=4-2\ep$, in case the integral diverges. If it is finite and $D=4$, the dual conformal transformation of the measure compensates exactly that of the propagator factors at points  $x_6,\ x_7$ and the integral is dual conformally covariant. With an appropriately  chosen numerator $N_{\rm pl}(p_i)$, depending on the external points only,  the integral becomes DCI.   If divergences are present, the weights of the measure and of the integrand do not match and the symmetry becomes anomalous. This can be formulated as a dual conformal Ward identity,
\begin{align}\label{1.3}
K^\mu\int  d^D x_6 d^D x_7\, {\cal I}_{\rm pl}(x_i;x_6, x_7) = 2\ep \int  d^D x_6 d^D x_7\,  (x^\mu_6 + x^\mu_7)\, {\cal I}_{\rm pl}(x_i;x_6, x_7)\,,
\end{align}
where ${\cal I}$ is the integrand as a function of the external and internal dual points, and 
\begin{align}\label{bK}
K^\mu = \sum_i (x^2_i \pa/\pa{x^\mu_i} - 2 x^\mu_i x^\nu_i \pa/\pa{x^\nu_i})
\end{align} 
is the generator of the special dual conformal transformations (boosts).

In the nonplanar case the above construction is not possible. The diagram on the right in Fig.~\ref{fig:dualcoords} explains why. We have added leg $p_3$ and pulled it out of the propagator frame, so that it takes its natural position between the dual points $x_3 \neq x_4$. By doing so, we have split the middle propagator into two. The bottom half is crossed by the dashed line connecting the integration points  $x_6,\ x_7$ as before, and it is represented by the propagator factor $1/x^2_{67}$. However, the top half corresponds to a dashed line between points $x_6,\ x_7$ that also crosses the new external line $p_3$. Examining momentum conservation, we see that this implies a shift of the top dual line by $p_3=x_{43}$. Its propagator factor becomes $1/(x_{67}+x_{43})^2$ and is not of the form $1/x_{ij}^2$ anymore. So, the DCI is lost for this nonplanar configuration.

The  key observation  of Refs.~\cite{Bern:2017gdk,Bern:2018oao} is that for this and similar nonplanar integrals one can preserve part of the dual conformal symmetry. It corresponds to projecting the boost generator with the shift parameter, in our case $p_3\cdot K$. Then one can show\footnote{See Appendix \ref{aA} for a detailed discussion of the dual conformal and directional  dual conformal symmetry.} that, if $p_3$ is lightlike, $p^2_3=0$, then
\begin{align}\label{1.5}
(p_3\cdot K)\, p^\mu_3=0\,.
\end{align}
This means that the infinitesimal transformations with parameter $\vep p^\mu_3$ (with $\vep\to0$) can be exponentiated and form a subgroup of the conformal group. More importantly, the shifted and unshifted propagators transform in exactly the same way, so that 
\begin{align}\label{}
(p_3\cdot K)\, \frac{(x_{67}+p_3)^2}{x^2_{67}} =0\,.
\end{align} 
We call this property {\it directional dual conformal invariance} (DDCI), meaning that the boost is projected on the lightlike direction of the external momentum $p_3$.

The integrand of the nonplanar diagram in Fig.~\ref{fig:dualcoords} has the general form
\begin{align}\label{}
{\cal I}_{\rm np}(x_i;x_6, x_7) = \frac{N_{\rm np}(x_i;x_6, x_7)}{x^2_{16}x^2_{26}x^2_{36} x^2_{67} (x_{67}+x_{43})^2 x^2_{37}x^2_{57}x^2_{17}}\,.
\end{align} 
It is clear that the denominator,  made of propagator factors, transforms covariantly under the DDC boosts. However, the conformal weights at the loop integration points $x_6\,, \, x_7$ have changed, compared to \p{2.2},  because of the new  shifted propagator. To match (or almost match in case of dimensional regularization) the weights of the measure, we need the numerator $N_{\rm np}$, which now depends on all the points. The rules how to construct numerators with the appropriate DDC weights are explained in Appendix \ref{aA}. Finally, we are in a position to formulate the DDCI Ward identity \cite{Bern:2017gdk,Bern:2018oao}
\begin{align}\label{1.7}
(x_{43} \cdot K) \int  d^D x_6 d^D x_7\, {\cal I}_{\rm np}(x_i;x_6, x_7) = 2\ep \int  d^D x_6 d^D x_7\  x_{43} \cdot (x_6 + x_7)\, {\cal I}_{\rm np}(x_i;x_6, x_7)\,,
\end{align}
or equivalently in terms of the momenta,
\begin{align}\label{2.9}
(p_3 \cdot K) \int  d^{D} \ell_{1} d^{D} {\ell_{2}} \, \mathcal{I}_{\rm np}( p_i;\ell_1, \ell_2  )  = 2\eps  \, \int  d^{D} {\ell_{1}} d^{D} {\ell_{2}} \ p_3\cdot (\ell_1 + \ell_2) \, \mathcal{I}_{\rm np}( p_i;\ell_1, \ell_2 )\,.
\end{align}

Our discussion easily generalizes to more legs and loops. However, one should bear in mind that the trick of preserving part of the DCI can only work for nonplanar graphs that can be reduced to planar ones by removing a single external leg. The reason for this is the key property \p{1.5}: it will not work for more than one projection of the boost generator $K_\mu$. We call this class of graphs `next-to-planar'. This may seem a rather restricted class but in fact it is not. Indeed, in Ref.~\cite{Bern:2018oao} it was shown   that all the two-loop four- and five-leg integrals of the ${\cal N}=4$ sYM amplitude are DDCI, each with its  appropriate projection $(p_i\cdot K)$. 
The list of DDCI integrals in \cite{Bern:2018oao} is not exhaustive, here we show many more examples. 

We remark that equation \p{2.9} resembles the canonical differential equations that are expected for any pure function \cite{Henn:2013pwa}.
Traditionally, differential equations are derived using so-called integration-by-parts identities, which at present require considerable amount of algebra.
One difference is that here, the fact that the r.h.s. is proportional to $\eps$ can be seen immediately, as it follows from the covariance of the integrand under the directional dual conformal symmetry. Therefore one can envisage using this equation also in situations that are not yet within the reach of standard IBP methods.

The main subject of this paper are the consequences of the (anomalous) Ward identity \p{2.9} for the integrated quantities.
In order to extract useful information from it, one can consider the Laurent expansion in $\eps$ of both sides of the equation. It is important to realize that by construction, due to the dual conformal power counting, the r.h.s. is ultraviolet finite, and stays finite even with one inserted loop momentum. On the other hand, the insertion on the r.h.s. does not worsen the infrared properties of the integral. From this we conclude that the integrals on the r.h.s. have no worse divergences than the integral on the l.h.s.. Thanks to the additional presence of a factor of $\eps$, we need to know the Laurent expansion of the r.h.s. to one order lower than then l.h.s.

In particular, it follows that the leading pole of the integrals satisfying (\ref{2.9}) will be invariant under $(p_3 \cdot K)$. Often, leading poles of Feynman integrals are simple constants, so one might wonder whether this is a trivial statement. Indeed, for example, the nonplanar four-point integral in Fig.~1 in Ref.~\cite{Bern:2018oao} has this property. However, here we will see that this is not always the case. 
It goes without saying that this argument also includes the case of finite integrals, which are invariant under the symmetry. This was already mentioned in \cite{Bern:2017gdk,Bern:2018oao}, but no explicit examples were presented. 

In Section \ref{s32} and Appendix \ref{sec:six-dim-examples}, we provide examples of DDCI integrals, both divergent and finite, and show that Eq.~(\ref{2.9}) holds for the first term in their Laurent expansion. In particular, we consider two-loop integrals whose leading poles are of lower degree than the typical $1/\ep^4$, and therefore their coefficients are nontrivial functions  satisfying the DDCI Ward identity.

Furthermore, in Section \ref{s34} we show how to use Eq.~(\ref{2.9}) to constrain subleading terms in the Laurent expansion efficiently. 
To this end, we use insights on the origin of the divergences of the integrals, in this way simplifying the calculation of the anomaly term on the r.h.s. of Eq.~(\ref{2.9}) considerably. This is very similar in spirit to the recent applications of conformal and superconformal symmetry to Feynman integrals \cite{Chicherin:2017bxc,Chicherin:2018ubl}.

\section{Implications of DDCI for pentagon functions}
\label{sec:implications-pentagon-functions}

In this section we analyze the implications of DDCI for five-particle integrals. 
Based on the knowledge of the one-loop and two-loop master integrals \cite{Gehrmann:2015bfy} in $D=4-2\ep$, in \cite{Chicherin:2017dob}
it was conjectured that the five-point massless diagrams evaluate to pentagon functions  characterized by a 31-letter alphabet. 
We specify the DDCI subspace in the space of the pentagon functions. Then we provide several examples of two-loop integrals whose leading poles in the $\ep$-expansion live in this subspace, and also check that the subleading terms satisfy the anomalous Ward identity \p{2.9}.
In Appendix~\ref{sec:six-dim-examples} we consider several six-dimensional nonplanar Feynman integrals which are IR finite, and we show that their integrated expressions are exactly DDCI.

\subsection{Five-particle DDCI variables}

Let us first consider the homogeneous Ward identity, 
\begin{align}\label{eq-bonus-homogeneous}
(p_{3} \cdot K) I = 0\,,
\end{align}
i.e. the case where the r.h.s. of Eq.~(\ref{2.9}) can be neglected. This applies to finite integrals or to the leading pole of a divergent integral, see Section~\ref{s32}.
In the following we assume that $I$ is dimensionless.

For on-shell five-particle scattering, $I$ can in general depend on four dimensionless variables.
The latter can be chosen, e.g., as ratios of the Mandelstam variables,  $\frac{s_{12}}{s_{15}},\frac{s_{23}}{s_{15}},\frac{s_{34}}{s_{15}},
\frac{s_{45}}{s_{15}}$.
The single condition Eq.~(\ref{eq-bonus-homogeneous}) can be used to remove one of the variables.
The analysis in Appendix \ref{a42} suggests several natural choices for the three independent invariants of the symmetry.
One choice is\footnote{They correspond to $\hat{u}_{13}, \; \hat{u}_{25}$ in Eq.~\p{115} and $\hat u_{1524}$ in Eq.~\p{ea27}.}
\begin{align}
\frac{s_{45}}{s_{12}} \;,\; \frac{s_{24}}{s_{15}} \;,\; \frac{s_{35} s_{23}}{s_{45}s_{24}}\,.
\end{align}
The Mandelstam invariants are Lorentz scalars, but the five-particle kinematics allows for nontrivial Lorentz pseudoscalars. 
Then it is also natural to build chiral DDCI variables out of the helicity spinors,
\begin{align} \label{xyz}
x = \frac{[24]\vev{45}[51]\vev{12}}{\vev{24}[45]\vev{51}[12]}\,, \quad y  = \frac{[35]\vev{51}[12]\vev{23}}{\vev{35}[51]\vev{12}[23]}\,, \quad z   = \frac{[13]\vev{34}[45]\vev{51}}{\vev{13}[34]\vev{45}[51]}\,.
\end{align}
In what follows we prefer the latter choice. So, the general solution to the homogeneous equation~(\ref{eq-bonus-homogeneous}) is
\begin{align}
I = I( x, y, z) \,.
\end{align}

In comparison to this, the full planar dual conformal symmetry, $K_\mu I = 0$ is much stronger. It eliminates four kinematic invariants. So, for a five-particle kinematics this symmetry only leaves the trivial solution $I = {\rm const}$.

\subsection{Constraints on the pentagon alphabet from DDCI}

We now argue that we can derive stronger consequences of the DDCI if 
we combine it with the recently acquired knowledge of the space of functions appearing in the
solution.

Let us begin with a brief review. Many classes of multi-point integrals, relevant for phenomenology, evaluate to hyperlogarithms, also called Goncharov polylogarithms.
These are iterated integrals, with arguments depending on the dimensionless kinematic variables of the problem. The massless five-particle scattering is of this type. Instead of working directly with these multi-variable functions, which have an intricate branch cut structure and satisfy numerous functional relations, nowadays it is common to consider their symbols. These are algebraic objects that reflect the iterated integral structure of the hyperlogarithms, but they lack the analytic information about the integration contour.

The arguments of the hyperlogarithm functions representing a given class of Feynman integrals can be characterized by an alphabet. 
Here we have in mind the pentagon alphabet  \cite{Chicherin:2017dob} of $31$ letters $W_{i}, i = 1\ldots 31$, which are functions of the kinematic variables of the problem (for their definitions see Appendix \ref{a42}). A symbol of weight $w$ is a linear combination of $w$-fold tensor products of the alphabet letters, which we denote by square brackets in what follows, $\sum_{i_1,\ldots,i_w} c_{i_1 \ldots i_w} [W_{i_1},\ldots,W_{i_{w}}]$. 
The symbols satisfy the logarithmic additivity property with respect to each tensor factor,  
\begin{align} \label{add}
[\ldots,W_{i}W_{j},\ldots] = [\ldots,W_{i},\ldots] + [\ldots,W_{j},\ldots]\,.
\end{align} 
It takes a more transparent form if we replace the tensor factors by their logarithms, but we prefer not to do it for the sake of brevity.
Each hyperlogarithm is represented by its symbol satisfying an integrability condition.
Derivatives act on the symbol's last entry, $\pa_v [W_{i_1},\ldots,W_{i_w}] = [W_{i_1},\ldots,W_{i_{w-1}}] \pa_{v} \log W_{i_w}$, reducing the weight by $1$. This reflects the differentiation formula for the corresponding iterated integral. Thus if a given hyperlogarithmic function satisfies a differential equation, e.g. the Ward identity \p{2.9}, so does its symbol. 

Let us consider functions in the space of the pentagon alphabet. The question then is which combinations of these letters are invariant under $(p_{3} \cdot K)$.
We find ten solutions,  
\begin{align}\label{10lettersW}
\{\alpha_i\}_{i=1}^{10}=\left\{ W_{26} \,,\, W_{27} \,,\, W_{30} \,,\, \frac{W_5}{W_{17}}\,,\, \frac{W_{10}}{W_{17}} \,,\, \frac{W_3 W_{16}}{W_{31}} \,,\, 
\frac{W_1 W_{17}}{W_{31}} \,,\, \frac{W_4 W_{5}}{W_{31}} \,,\, \frac{W_{11} W_{17}}{W_{31}} \,,\, \frac{W_2 W_{18}}{W_{31}} \right\}
\end{align}
The first three letters $\alpha_1,\alpha_2,\alpha_3$ are parity odd (i.e. $\log \alpha_i$, $i=1,2,3$, changes sign under parity)  and the remaining seven letters are parity even. 
As we discussed above, the ten letters are functions of three independent DDCI variables. 
The first three letters in \p{10lettersW} coincide with the variables defined in Eq. \p{xyz}, i.e. $x=W_{26},\, y =W_{27} ,\, z= W_{30}$.
Then the ten solutions \p{10lettersW} are functions of $x,y,z$,
\begin{align}
& \alpha _1=x\, ,\; \alpha _2=y \, ,\;\alpha _3=z \,,\;\alpha _4=-\frac{(1-x y) (1-x z)}{x (1-y) (1-z)}\,,\; 
\alpha _5=-\frac{(1-x) (1-x y z)}{x (1-y) (1-z)}\,, \nt
&	\alpha _6=\frac{(1-x)z}{(1-z) (1-x z)}\,,\; \alpha _7= \frac{x (1-z)}{(1-x) (1-x z)}\,,\; 
\alpha _8=-\frac{(1-x z)}{(1-x) (1-z)}\,,\nt 
&\alpha _9=\frac{x (y-z)}{( 1- x y) ( 1- x z)}\,,\;\alpha _{10}=\frac{(1-x) y}{(1-y) (1- x y)}\,. 
\end{align}
In view of the logarithmic additivity \p{add} the ten-letter alphabet can be equivalently chosen as follows 
\begin{align}\label{10lettersxyz}
\{ x \; , \;  1-x \;, \; y  \;, \; 1-y \;,\; z \;, \; 1-z \;, \; 1- x y \;, \; 1- x z \;,\; y - z \;, \; 1- x y z  \} \,.
\end{align}
The first entry of the symbol representing a Feynman graph is related to its discontinuities. 
For  massless scattering only the Mandelstam variables $s_{ij}$ are allowed first entries. 
Thus only the following four combinations of the letters $\{\alpha_i\}$ can serve as first entries in the framework of the ten-letter alphabet (\ref{10lettersW}),
\begin{align}\label{firstenties10lettersW}
\text{First entries for the $\{\a\}$ alphabet:} \qquad \frac{W_1}{W_{4}}\;,\; \frac{W_5}{W_{17}} \;, \; \frac{W_3 W_{16}}{W_2 W_{18}} \;,\; \frac{W_1 W_{17}}{W_2 W_{18}} \,.
\end{align}

\begin{table}
\begin{center}
\begin{tabular}{l|c|c|c|c}
weight & 1 & 2 & 3 & 4 \\ \hline
alphabet $\{\a\}$ & $4|0$ & $13|2$ &  $44|12$ & $148|62$ \\ \hline 
alphabet $\{\b\}$ & $2|0$ & $3|1$ & $6|3$ & $12|9$ \\ \hline
alphabet $\{\gamma\}$ & $1|0$ & $2|0$ & $4|0$ & $8|0$
\end{tabular}
\end{center}
\caption{Number of parity even$|$odd integrable symbols up to weight four for the ten-letter $\{\alpha\}$ \p{10lettersW}, five-letter $\{\beta\}$ \p{5letters}, and two-letter  $\{\gamma\}$  alphabets satisfying the first entry conditions.}
\label{tabsymb}
\end{table}

We note that integrals with an enhanced permutation symmetry, such as topology (i) in Section~\ref{s322}, which is invariant under the permutations of $p_2,p_3,p_5$, 
will satisfy three DDCI relations. Let us consider the consequences of this symmetry. Introducing dual coordinates for each choice of the nonplanar loop momentum ($p_2, p_3$ or $p_5$), we find that two of these three relations are independent. There are five solutions
\begin{align}
\{ \beta_i\}_{i=1}^{5} =\left\{ W_{26} \;,\; W_{30} \;,\; \frac{W_3 W_{16}}{W_{31}} \;,\; 
\frac{W_1 W_{17}}{W_{31}} \;,\; \frac{W_4 W_{5}}{W_{31}} \right\} \,, \label{5letters}
\end{align}
which is a subset of the ten-letter alphabet, $\{\beta_i\} \subset\{ \alpha_i\}$. The five-letter alphabet allows
only two first entries, 
\begin{align}
\text{First entries for the $\{\b\}$ alphabet:} \qquad \frac{W_4 W_5}{W_{1} W_{17}}\;,\; \frac{W_3 W_{16}}{W_{1} W_{17}} \,.
\end{align}
We remark that the five-letter alphabet (\ref{5letters}) can be equivalently expressed in terms of the following letters,
\begin{align}
\{ x \; , \;  1-x \;, \; z  \;, \; 1-z \;,\; 1- x z   \} \,.
\end{align}

Finally, we mention that there exist {\it planar} integrals that are DDCI, but not DCI. In this case, the propagator factors in the denominator are not shifted and are dual-conformally covariant,
but one can construct numerators that are covariant under the directional symmetry only. Recall that a finite DCI five-particle integral is necessarily a constant, as 
there are no invariants. In the DDCI case, the starting point is (\ref{10lettersW}), with the additional restriction that the first entry should be drawn from the set $\{ W_{1}, W_{2}, W_{3}, W_{4}, W_{5}\}$.  Comparing to Eq.~(\ref{firstenties10lettersW}), we see that only the variable $w=W_{1}/W_{4} $ fulfills this requirement. As a consequence, planar DDCI functions are given by the alphabet 
\begin{align}
\{ \gamma_{i} \}_{i=1}^{2} = \left\{ \frac{W_{1}}{W_{4}}  , \frac{W_{4}}{W_{11}} \right\} = \left\{ w, -1+w \right\} \,.
\end{align}

The number of integrable symbols for the three alphabets up to weight four is summarized in Tab.~\ref{tabsymb}.

Remarkably, the various alphabets that we encountered in this section are relatively simple. They all belong to a class of alphabets related to the moduli space of $n$ marked points on a sphere \cite{Brown:2009qja}, with $n=6,5,4$, respectively. All of these alphabets have appeared previously in physics applications. Here we mention especially the case $n=6$, which appears in planar six-particle scattering in $\mathcal{N}=4$ sYM \cite{Goncharov:2010jf}.\footnote{Strictly speaking, for dual conformal six-particle scattering amplitudes, only $9$ combinations of the $10$ letters are needed. It is easy to check that in the examples of DDCI pentagon functions studied here, likewise at most $9$ letters appeared, namely combinations without $W_{31}$. \label{footnote9vs10}}

\subsection{Checking the DDCI Ward identity for the leading poles: Weight two and three functions}\label{s32}

Here we consider several two-loop integrals with DDCI integrands in $D=4$. The dimensional regularization with $D=4-2\ep$, employed for treating the IR divergences, breaks the naive symmetry. The examples in this subsection allow us to verify that the functional expressions for the integrals  satisfy the anomalous Ward identity  \p{2.9} at the level of the leading pole in $\ep$. In this way we provide evidence for the DDCI of the integrals, which extends the naive invariance of their integrands.

\begin{figure}
\begin{center}
\includegraphics[width = 6cm]{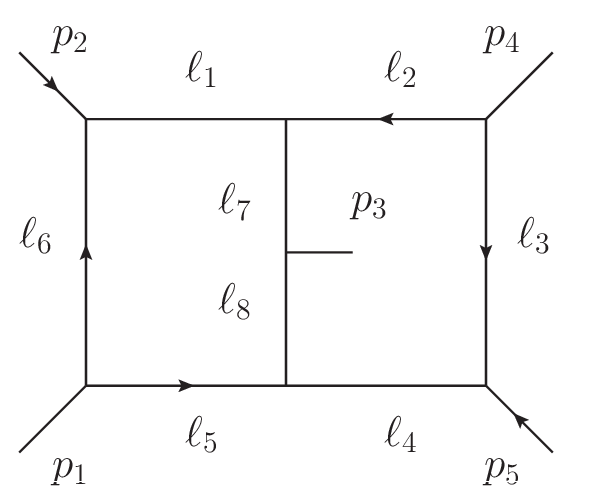}
\qquad\qquad
\includegraphics[width = 6cm]{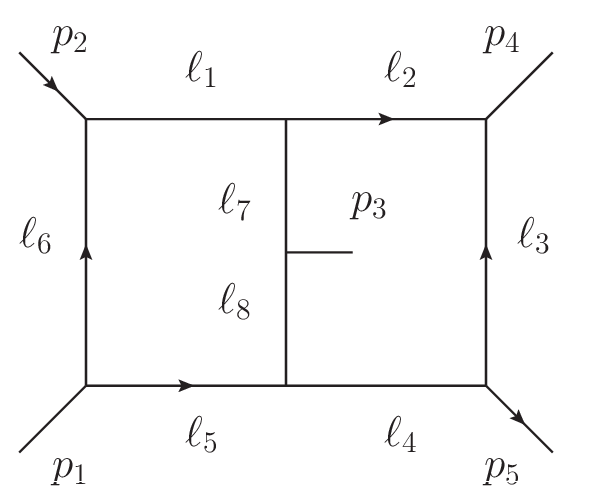}
\end{center}
\caption{Integrals of topology $(a)$ with numerators $N_{a_1}$ (left) and $N_{a_2}$ (right).  The arrows depict the fermion propagators and fermion external states that form the numerators.} \label{topAint}
\end{figure}

\subsubsection{Topology (a)}

We consider the nonplanar five-point integrals shown in Fig.~\ref{topAint}, 
\begin{align} \label{intA}
I_{a_k} = \int \frac{ d^D \ell_1 d^D \ell_2}{(i\pi^{D/2})^2}\frac{N_{a_k}}{\ell_1^2 \ldots \ell_8^2}\,,  \quad k=1,2,
\end{align}
with numerators 
\begin{align}
N_{a_1} = \vev{2|\ell_6 \tilde\ell_5|3} \vev{3|\ell_2 \tilde\ell_3|5} \;, \qquad
N_{a_2}=\vev{2|\ell_6 \tilde\ell_5|3} [3|\ell_2 \tilde\ell_3|5]\,.
\end{align}
It is easy to check that both integrands in \p{intA} are DDC covariant in four dimensions.

These integrals are of topology (a) according to the classification in Ref.~\cite{Bern:2018oao},
but their numerators differ from those of the integrals contributing to the $\cN = 4$ sYM five-particle amplitude \cite{Bern:2015ple}.  The reason for this choice is that the integrals in Fig.~\ref{topAint} have an improved IR behavior,
\begin{align} \label{ep-exp}
I_{a_k} = \frac{1}{\ep} \, I^{(3)}_{a_k} +  \ep^0 \, I^{(4)}_{a_k} + {\cal O}(\ep)\,,
\end{align}
where the functions $I^{(w)}_{a_k}$ are of transcendental weight $w$. This enables us to present higher-weight examples of DDCI.

For these integrals the only source of IR divergences is the regime in which the loop momentum $\ell_7$ becomes collinear with the on-shell momentum $p_3$.
This region of the loop integration is responsible for the $\frac{1}{\ep}$-term in \p{ep-exp}.  In order to extract the pole we combine the propagators $\ell_7$ and $\ell_8$ by introducing a Feynman parameter. Then we pick out the singular (contact) term of the resulting distribution,  
\begin{align} 
\frac{1}{\ell_7^2 \ell_8^2} = \int^1_0 d\xi \, \frac{1}{(\ell_7 + \xi p_3)^4}\ \to\  \frac{i \pi^2}{\ep} \int^1_0 d\xi \, \delta^{(4)}(\ell_7 + \xi p_3)\,. \label{gelf}
\end{align}
In this way we cut  the middle propagator in the left diagram  in Fig.~\ref{topAint}, and we
find that the residue of the pole $I_{a_1}^{(3)}$ is given by a one-fold integral of a hexagon with numerator $N_{a_1}$,
\begin{align} 
I^{(3)}_{a_1} = \int^1_0 d\xi \begin{array}{c}\includegraphics[width=5cm]{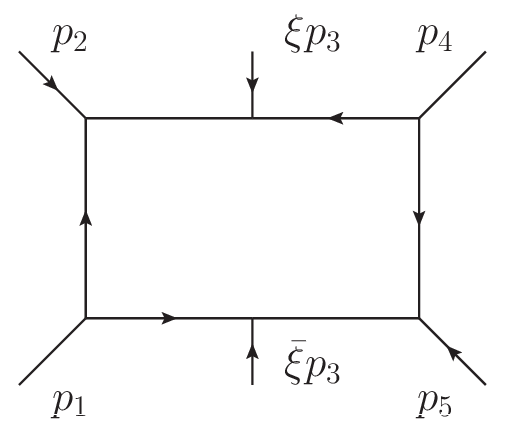}\end{array}
\end{align} 
The arrows depict the fermionic propagators $\ell_{\a\da}/\ell^2$ and fermionic external states $\ket{i}$, which are arranged in a way to form the numerator $N_{a_1}$.
This one-loop integral is the `magic' hexagon with chiral numerator considered in \cite{ArkaniHamed:2010gh}. 
The hexagon is dual conformal in a generic six-particle kinematics. Its explicit expression was also given in \cite{ArkaniHamed:2010gh}. So we obtain a very simple representation for the leading pole term of the integral $I_{a_1}$,
\begin{align} 
I^{(3)}_{a_1} = \frac{1}{[32][35]}\int^1_0 \frac{d\xi}{\xi \bar\xi} \left( {\rm Li}_2(1-u_1) + {\rm Li}_2(1-u_2)  + {\rm Li}_2(1-u_3) + \log u_1 \log u_3 - \frac{\pi^2}{3}\right) \,.\label{IaPole}
\end{align}
Here $u_1,u_2,u_3$ are the dual conformal cross-ratios for the six-particle kinematics, 
\begin{align}\label{3.14}
&u_1 = \frac{s_{12} s_{45}}{(\bar\xi s_{12} + \xi s_{45}) (\xi s_{12} + \bar\xi s_{45})} \,, \nt
&u_2 = \frac{\xi \bar\xi s_{23} s_{35}}{(\bar\xi s_{12} + \xi s_{45})(\bar\xi s_{15} + \xi s_{24})} \,, \nt
&u_3 = \frac{\xi \bar\xi s_{13} s_{34}}{(\xi s_{12} + \bar\xi s_{45})(\xi s_{15} + \bar\xi s_{24})}\,,
\end{align}
restricted to the five-particle configuration, and $\bar\xi \equiv 1-\xi$. In this form DDCI is not manifest. We explain how one can show it in Appendix \ref{appA41}.

Evaluating the one-fold integral in \p{IaPole} we find
\begin{align}\label{321}
I^{(3)}_{a_1} = \frac{1}{[32][35]} {\cal P}_1\,.
\end{align}
Here the rational prefactor agrees with the  analysis of the leading singularities of the integral $I_{a_1}$, and ${\cal P}_1$ is a pure weight-three function whose symbol is given by
\begin{align}\label{resultP1}
& {\cal P}_1 =   \frac{1}{2}
   \left[\frac{W_1 W_{17}}{W_2 W_{18}},\frac{W_1^2 W_3 W_5 W_{16} W_{26}
   W_{27}}{W_2 W_4^2 W_{17} W_{18} W_{30}},W_{26}\right]+\frac{1}{2}
   \left[\frac{W_4 W_{17}}{W_2 W_{18}},\frac{W_{11}^2 W_{17}}{W_1 W_3
   W_{16}},\frac{W_4 W_5 W_{26}}{W_1 W_{17}}\right]    \nt & 
	 + 2 \left[\frac{W_1}{W_4},\frac{W_1}{W_4},\frac{W_5 W_{11}^2}{W_3 W_4
   W_{16}}\right]+\frac{1}{2} \left[\frac{W_5}{W_{17}},\frac{W_3 W_{16}
   W_{17}}{W_1 W_{10}^2},\frac{W_4 W_5 W_{26}}{W_1 W_{17}}\right]   +\left[\frac{W_3 W_{16}}{W_2
   W_{18}},\frac{W_1}{W_4},\frac{W_4 W_{17}}{W_1 W_5 W_{26}}\right] \nt &
   +\frac{1}{2} \left[\frac{W_4
   W_{17}}{W_2 W_{18}},\frac{W_2 W_4^2 W_5 W_{11}^2 W_{18}}{W_1^3 W_3^2
   W_{16}^2},\frac{W_4 W_{17}}{W_1 W_5 W_{26}}\right]+\frac{1}{2}
   \left[\frac{W_4 W_{17}}{W_2 W_{18}},\frac{W_2 W_4^2 W_5 W_{18} W_{26}
   W_{30}}{W_1^2 W_3 W_{16} W_{17} W_{27}},W_{26}\right]    \nt &	 
	+\frac{1}{2}
   \left[\frac{W_1 W_{17}}{W_2 W_{18}},\frac{W_3 W_4 W_{16}}{W_5
   W_{11}^2},\frac{W_4 W_5 W_{26}}{W_1 W_{17}}\right]+\frac{1}{2}
   \left[\frac{W_1 W_{17}}{W_2 W_{18}},\frac{W_1^2 W_3^2 W_{16}^2}{W_2 W_4
   W_{11}^2 W_{17} W_{18}},\frac{W_4 W_{17}}{W_1 W_5 W_{26}}\right]   \nt &  +\frac{1}{2}
   \left[\frac{W_5}{W_{17}},\frac{W_4 W_{10}^2}{W_2 W_{17} W_{18}},\frac{W_4
   W_{17}}{W_1 W_5 W_{26}}\right]+\frac{1}{2} \left[\frac{W_5}{W_{17}},\frac{W_1
   W_4 W_{10}^4 W_{26}^2 W_{27} W_{30}}{W_2 W_3 W_{16} W_{17}^2
   W_{18}},W_{26}\right] \nt &
 +\frac{1}{2}
   \left[\frac{W_3 W_{16}}{W_2 W_{18}},\frac{W_1 W_5}{W_4 W_{17}},\frac{W_4 W_5
   W_{26}}{W_1 W_{17}}\right]+\frac{1}{2} \left[\frac{W_3 W_{16}}{W_2
   W_{18}},\frac{W_1 W_{17}}{W_4 W_5 W_{26}},W_{26}\right]\,.
\end{align}
We observe that ${\cal P}_1$   depends only on the $10$ letters in Eq.~(\ref{10lettersW}). Equivalently, it can be written in terms of the $10$ letters  (\ref{10lettersxyz}), and is a function of
$x,y,z$ only.
So we see that  the Ward identity $(p_3\cdot K) {\cal P}_1 = 0$ is satisfied, as expected.\footnote{The rational prefactor in \p{321} is DDC covariant, see \p{b17}.}

The residue $I^{(3)}_{a_2}$ of the second integral is obtained in a similar manner. Extracting the pole of $I_{a_2}$ with the help of \p{gelf},  
we represent $I^{(3)}_{a_2}$ as a one-fold integral of a hexagon with numerator $N_{a_2}$. It is the other `magic' hexagon from \cite{ArkaniHamed:2010gh} 
with mixed chiral-antichiral numerator. Substituting its explicit expression we obtain   
\begin{align} 
I^{(3)}_{a_2} = \bra{2} p_4|5] \int^1_0 d\xi \biggl( & - \frac{s_{12}}{\xi\bar\xi(s_{12}-s_{45})} \frac{\log u_1 \log u_2}{(s_{12}s_{24}+t_{1342}\,\xi)}+ \frac{\xi s_{12} +\bar\xi s_{45}}{\xi\bar\xi(s_{12}-s_{45})} \frac{\log u_1 \log u_3}{(s_{24}s_{45} + t_{5324}\, \xi)} \nt 
&+ \frac{s_{13}s_{34} \log u_2 \log u_3}{(s_{13}s_{34}\xi+ t_{1243})(s_{24}s_{45}+t_{5324}\,\xi)} \biggr) \,,\label{IbPole}
\end{align}
where $t_{ijkl} \equiv \vev{ij}[jk]\vev{kl}[li]$. This expression is DDCI, as explained in Appendix~\ref{appA41}. Implementing the $\xi$-integration we find
\begin{align}\label{3.17}
I^{(3)}_{a_2} = - \frac{\vev{13}[34]}{\vev{15}[24](s_{12}-s_{45})} {\cal P}_2 
+ \frac{\vev{12}}{\vev{15}(s_{12}-s_{45})} {\cal P}_3 \,.
\end{align}
Here we observe  two leading singularities, both DDCI (see \p{b17}). The pure weight-three functions  ${\cal P}_2$ and ${\cal P}_3$ are represented by the following symbols
\begin{align}
& {\cal P}_2 = \frac{1}{2}
   \left[\frac{W_5}{W_{17}},\frac{W_1 W_2 W_4 W_{10}^4 W_{18}}{W_3^3 W_5^2
   W_{16}^3},\frac{W_1 W_2 W_{18}}{W_3 W_4 W_{16}}\right]  +\frac{1}{2} \left[\frac{W_2 W_3 W_{16}
   W_{18}}{W_1 W_4 W_5^2},\frac{W_2 W_{17} W_{18}}{W_3 W_5 W_{16}},\frac{W_1 W_2
   W_{18}}{W_3 W_4 W_{16}}\right]  \nt &
   +\left[\frac{W_1 W_{17}}{W_2 W_{18}},\frac{W_1 W_2 W_{18}}{W_3 W_4
   W_{16}},\frac{W_4}{W_1}\right]+\frac{1}{2} \left[\frac{W_1 W_{17}}{W_2
   W_{18}},\frac{W_2^2 W_{18}^2}{W_3^2 W_{16}^2},\frac{W_1 W_2 W_{18}}{W_3 W_4
   W_{16}}\right]   +\frac{1}{2}
   \left[\frac{W_5}{W_{17}},\frac{W_{27}}{W_{30}^3},\frac{W_{30}}{W_{27}}\right]  
	\nt &  -\frac{1}{2} \left[\frac{W_1 W_{17}}{W_2
   W_{18}},\frac{W_{30}^2}{W_{27}^2},\frac{W_{30}}{W_{27}}\right]+\left[\frac{W_
   2 W_4^2 W_{18}}{W_1^2 W_3 W_{16}},\frac{W_2 W_4 W_{18}}{W_1 W_3
   W_{16}},\frac{W_4}{W_1}\right]   +  \left[\frac{W_5}{W_{17}},\frac{W_1^2 W_{10}^2}{W_3^2
   W_{16}^2},\frac{W_4}{W_1}\right]  \nt &   +\frac{1}{2} \left[\frac{W_2 W_4^2
   W_{18}}{W_1^2 W_3 W_{16}},\frac{W_2 W_4 W_{18}}{W_1 W_3 W_{16}},\frac{W_1 W_2
   W_{18}}{W_3 W_4 W_{16}}\right]   +\left[\frac{W_2
   W_3 W_{16} W_{18}}{W_1 W_4 W_5^2},\frac{W_1 W_{17}}{W_3
   W_{16}},\frac{W_4}{W_1}\right]   \nt &   + \frac{1}{2} \left[\frac{W_2 W_3 W_{16}
   W_{18}}{W_1 W_4 W_5^2},\frac{W_{26} W_{27}}{W_{30}},\frac{W_{30}}{W_{27}}\right] 
	+\frac{1}{2} \left[\frac{W_2 W_4^2
   W_{18}}{W_1^2 W_3
   W_{16}},\frac{W_{27}}{W_{30}},\frac{W_{30}}{W_{27}}\right]  
\end{align}
and
\begin{align}
{\cal P}_3 = 2 \left[ \frac{W_1}{W_4},\frac{W_4}{W_1},\frac{W_2 W_{18}}{W_3 W_{16}}\right]
+2\left[\frac{W_1}{W_4},\frac{W_2 W_{18}}{W_3 W_{16}},\frac{W_4}{W_1}\right]+
2\left[\frac{W_3 W_{16}}{W_2 W_{18}},\frac{W_4}{W_1},\frac{W_4}{W_1}\right]\,.
\end{align}
They depend only on the reduced alphabet \p{10lettersW}, and hence the Ward identity for the leading term $I^{(3)}_{a_2}$ is satisfied.

\begin{figure}
\begin{center}
\includegraphics[width = 6cm]{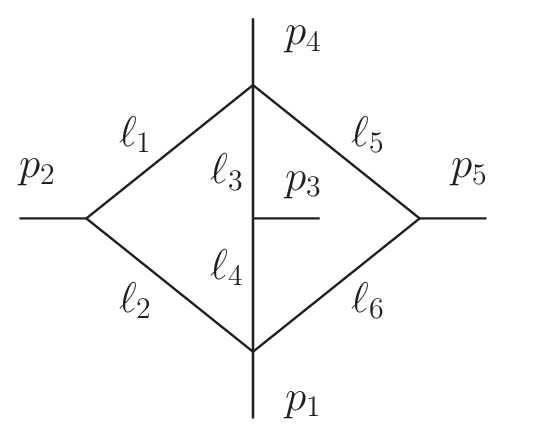}
\end{center}
\caption{Integral of topology (i)} \label{topi}
\end{figure}

\subsubsection{Topology (i)}\label{s322}

Our final example is the integral $I_{(i)}$ depicted in Fig.~\ref{topi}, 
\begin{align} \label{inti}
I_{(i)} = \int \frac{ d^D \ell_1 d^D \ell_2}{(i\pi^{D/2})^2} \frac{N_{(i)}}{\ell_1^2 \ldots \ell_6^2}\,.
\end{align}
It has the numerator $N_{(i)} = -4 i\ep(p_2,p_3,p_4,p_5) = \vev{23}[34]\vev{45}[52] - [23]\vev{34}[45]\vev{52}$, which is independent of the loop momenta. According to the table \p{b17}, the integrand of (i) is DDC covariant, as was already observed in \cite{Bern:2018oao}. This integral appears in the $\cN = 4$ sYM five-particle amplitude \cite{Bern:2015ple}.  Its $\ep$-expansion has the form
\begin{align}
I_{(i)} = \frac{1}{\ep^2}  I^{(2)}_{(i)} + \frac{1}{\ep}  I^{(3)}_{(i)} + \ep^0  I^{(4)}_{(i)} + {\cal O}(\ep) \,,\label{Iexp}
\end{align}
where the pole and finite terms are known \cite{Chicherin:2017dob}. Here and in the next subsection we  show that these nontrivial functions satisfy the DDCI  Ward identity \p{2.9}, not only at the level of the leading but also the subleading pole.  The latter is an example of the DDCI anomaly introduced by the IR divergences.

The integral \p{inti} diverges in the region where the loop momenta become collinear with the external momenta.
Each of the two loop momenta can be collinear with one of $p_2,p_3,p_5$. 
In order to extract the leading pole contribution to $I_{(i)}$ we apply the trick \p{gelf} twice. Choosing the collinear region specified by the momenta $p_2,p_3$, we combine  the propagators $\ell_1$ and $\ell_2$ giving a $1/\ep$ pole, and the propagators $\ell_3$ and $\ell_4$ giving another $1/\ep$ pole. Both loop integrations are localized and the remaining diagram has tree topology. We need to sum over the three possible choices of two momenta out of three, $\mathfrak{S} = \{(2,3),(2,5),(3,5)\}$, with the result   
\begin{align}
I^{(2)}_{(i)} = N_{(i)} \sum_{(n,m) \in \mathfrak{S}} \int^1_0 d\xi \int^1_0 d\eta \frac{1}{(s_{1 n} \xi + s_{1 m} \eta + \xi\eta s_{n m})(s_{4 n} \bar\xi + s_{4 m} \bar\eta + \bar\xi\bar\eta s_{n m})} \,. \label{topipole}
\end{align}
After the $\xi$ integration the three terms in \p{topipole} become identical and we find the leading pole in agreement with  \cite{Chicherin:2017dob}, 
\begin{align}\label{330}
I^{(2)}_{(i)} = 6 \left[ {\rm Li}_2(W_{26}) + {\rm Li}_2(W_{30}) - {\rm Li}_2(W_{26} W_{30}) - \frac{1}{2}\log(W_{26})\log(W_{30}) 
- \frac{\pi^2}{6}\right].
\end{align}
It belongs to the pentagon function subspace characterized by the five-letter alphabet \p{5letters}, and consequently it is DDCI.

\subsection{Implications of DDCI for the subleading poles}\label{s34}

Substituting the $\ep$-expansion \p{Iexp} in the anomalous Ward identity \p{1.7}, we expect that the subleading term satisfies the inhomogeneous equation
\begin{align}  \label{ianom}
\left(p_3 \cdot K\right) I^{(3)}_{(i)} =N_{(i)}\, \lim_{\ep \to 0} 4\ep^2 \int \frac{ d^D x_6 d^D x_7}{(i\pi^{D/2})^2} \frac{p_3\cdot(x_6 + x_7)}{x_{62}^2 x_{63}^2 x_{71}^2 x_{75}^2 x_{67}^2 (x_{67}+p_3)^2} \equiv (*) \,,
\end{align}
where we use the dual coordinates \p{1.1} for $n=5$.
In order to evaluate the r.h.s. of \p{ianom} we combine pairs of propagators and localize both loop integrations, as in the calculation of $I_{(i)}^{(2)}$. We obtain three contributions,
\begin{align}
(*) = 4 N_{(i)} \sum_{(n,m) \in \mathfrak{S}} \int^1_0 d\xi \int^1_0 d\eta \frac{V_{(n,m)}}{(s_{1 n} \xi + s_{1 m} \eta + \xi\eta s_{n m})(s_{4 n} \bar\xi + s_{4 m} \bar\eta + \bar\xi\bar\eta s_{n m})} \,,\label{xieta}
\end{align}
where we use the short-hand notations
\begin{align}
V_{(2,5)} = p_3 \cdot (x_1+x_3)-\frac12 ( \bar\xi s_{23} + \eta s_{35}) \;,\;
V_{(2,3)} = 2 p_3 \cdot x_3 - \bar\xi s_{23} \;,\;
V_{(3,5)} = 2 p_3 \cdot x_1 - \eta s_{35} \,.
\end{align}
Evaluating the two-fold integral in \p{xieta} we obtain the explicit expression for the r.h.s. of Eq.~\p{ianom} 
\begin{align} \label{anomExpl}
(*) = -\frac{6}{s_{25}} \left[ -4i \ep(p_1,p_2,p_3,p_4)\cdot T_1 + (s_{12} s_{35} - s_{15}s_{23} + s_{13}s_{25}) \cdot T_2 \right]\,,
\end{align}
where $T_1$ and $T_2$ are pure weight-two functions represented by the following symbols
\begin{align} 
& T_1 = \left[\frac{W_1}{W_3},\frac{W_{13} W_{23}}{W_4
   W_5}\right]+  \left[\frac{W_5}{W_3},\frac{W_1
   W_{17}}{W_9 W_{15}}\right]+  \left[\frac{W_3
   W_{16}}{W_1 W_{17}},\frac{W_1 W_{23}}{W_5
   W_9}\right]+  \left[\frac{W_1 W_{17}}{W_4
   W_5},\frac{W_1 W_{16}}{W_3 W_9}\right] ,
	\nt
& T_2 = \left[\frac{W_3 W_{16}}{W_1 W_{17}},W_{26}\right]+ \left[\frac{W_1 W_{17}}{W_4 W_5},W_{26} W_{30}\right] . 	
\end{align}
One can easily see that the anomaly involves more than the ten letters \p{10lettersW}. Indeed, we do not expect that the anomaly itself be DDCI.	The symbol $T_1$ is parity even, and $T_2$ is parity odd, and the whole expression \p{anomExpl} is parity odd due to the pseudoscalar factor $4i\ep(p_1,p_2,p_3,p_4)$. 
This is consistent with the fact that the integral (i) itself is parity odd.
The explicit expression for the symbol of $I^{(3)}_{(i)}$ is known \cite{Chicherin:2017dob}. It belongs to the full 31-letter pentagon space. We have checked that the variation $(p_3 \cdot K)$ of the symbol of $I^{(3)}_{(i)}$ coincides with Eq.~\p{anomExpl}. Thus we have explicitly verified that the Ward identity \p{ianom} is satisfied. 

The subleading pole of the integral (i) allowed us to probe the inhomogeneous Ward identity \p{2.9}.
This is a very nontrivial check, demonstrating the implications of directional dual conformal symmetry for IR divergent integrals, which are not exactly DDCI.  

In App.~\ref{sec:six-dim-examples} we provide further examples of the Ward identity \p{2.9} at work. 
We consider finite 6D integrals, and demonstrate that the DDCI of their integrands implies the exact DDCI invariance of the integrated expressions.

\section{Discussion and outlook}
\label{sec:discussion}

The restriction of dual conformal symmetry to one projection of the special conformal transformation is relatively weak, removing one variable only. 
On the other hand, we argued that, with some additional information the symmetry can be used effectively to constrain the possible function space.
Starting from the pentagon alphabet \cite{Gehrmann:2015bfy,Chicherin:2017dob}, we showed that the symmetry 
reduces the latter drastically, namely from $31$ to $10$ letters. 

There are several remarkable features of the $10$-letter alphabet we found:
\begin{itemize}
\item It is a very well-known alphabet, corresponding to six marked points on the sphere \cite{Brown:2009qja}.
Moreover, the same alphabet also describes (conjecturally) planar dual conformal six-particle scattering amplitudes \cite{Goncharov:2010jf}.\footnote{Cf. footnote \ref{footnote9vs10}.}
\item It is a subset of the {\it planar} pentagon alphabet from Ref.~\cite{Gehrmann:2015bfy}
\end{itemize}
For both of these observations, one should keep in mind that while the alphabets are related to those two cases, the specific functions that can appear (e.g. in a classification of integrable symbols) are different. The reason for this is that the first-entry conditions differ.
Nevertheless, the two observations suggest to us an intriguing simplicity of DDCI pentagon functions. As we discuss below, this is also relevant for constraining pentagon remainder functions. 
It would be interesting to investigate the cluster algebra properties of nonplanar dual conformal integrals, 
similar to the planar case \cite{Golden:2013xva,Parker:2015cia}.

Having a much smaller alphabet, and with nice algebraic properties, is a huge simplification.
Generic five-particle integrals were already successfully bootstrapped within the $31$-letter alphabet \cite{Gehrmann:2015bfy,Chicherin:2017dob}. 
Our result suggests that this method can be particularly powerful when applied to DDCI integrals.

We argued that for divergent integrals, the leading pole should be exactly DDCI. One might think that the leading pole of an integral is rather trivial, but this is not always the case. We demonstrated the invariance explicitly in a number of non-trivial cases, where the leading pole is given by weight two and weight three functions. As predicted, the leading pole is expressed in terms of the alphabet (\ref{10lettersxyz}).

Moreover, we showed how to use the Ward identity to constraint the subleading poles of integrals. In order to achieve this, we used knowledge of the origin of the collinear divergences, which allowed us to determine the inhomogeneous term of the DDC Ward identity. Using a known result from Ref.~\cite{Chicherin:2017dob}, we verified this latter identity successfully.

Furthermore, in Appendix \ref{sec:six-dim-examples} we presented examples of finite DDCI integrals. They include cases where in addition to the DDC symmetry, also the original conformal symmetry is present. Understanding the implications of the latter for on-shell integrals is a difficult question, but recent progress has been made \cite{Chicherin:2017bxc,Chicherin:2018ubl}. It is enticing to think about `Yangian' invariant objects having both symmetries, in analogy with the planar case \cite{Henn:2011xk,Chicherin:2017frs}.

All integrals appearing in the two-loop five-particle ${\cal N}=4$ sYM amplitude are DDCI with respect to some external momentum  \cite{Bern:2018oao}. As different integrals in the amplitude are invariant with respect to DDC generators projected along distinct legs, it is natural to decompose the amplitude according to $\mathcal{A} = \sum_{i=1}^{5} \mathcal{A}_{i}$, with each partial amplitude $\mathcal{A}_{i}$ annihilated by the DDC generator projected along $p_{i}$.
Of course, this invariance is only formal, due to infrared divergences.

In the planar case, these divergences, as well as how exactly they break the DCI, were understood thanks to the  duality with Wilson loops. In this way, an all-orders dual conformal Ward identity was formulated. The amplitude could then be expressed as a particular solution to that identity, plus a remainder function \cite{Drummond:2007au} (and, in the case of non-MHV amplitudes, ratio function \cite{Drummond:2008vq}) that is exactly DCI.
It will be very important to investigate whether a similar understanding can be achieved in the present context. See Refs.~\cite{Almelid:2015jia,Henn:2016jdu} for the current status of infrared divergences for nonplanar scattering amplitudes, and Ref.~\cite{Ben-Israel:2018ckc} for attempts at generalizing the scattering amplitudes / Wilson loop duality beyond the planar limit.

While the most suitable definition of the partial amplitudes $\mathcal{A}_{i}$, and the precise form of an all-order Ward identity remain to be discovered, we can already anticipate to what extent such an equation can fix the answer. Our results show that any such remainder function can depend on three variables $x,y,z$ only, see Eq. (\ref{xyz}). Moreover, together with the conjectured pentagon function space \cite{Chicherin:2017dob}, only the small subset (\ref{10lettersxyz}) of alphabet letters needs to be considered.


\section*{Acknowledgements}
This research received funding in part from a GFK fellowship and from  the PRISMA Cluster of Excellence at Mainz university, and from the European Research Council (ERC) under the European Union's Horizon 2020 research and innovation programme (grant agreement No 725110), {\it{Novel structures in scattering amplitudes}}.
\appendix

\section{Dual conformal symmetry}\label{aA}

In this Appendix we summarize the necessary information about dual conformal symmetry and its directional version. We analyze various ways of constructing invariants. 

\subsection{Spinor conventions}\label{aA1} 

We use the two-component spinor conventions of Ref.~\cite{Galperin:1984av}. They include the definitions of the Levi-Civita tensors
\begin{align}\label{}
\ep_{12}=-\ep^{12}=
\ep_{\dot{1}\dot{2}}=-\ep^{\dot{1}\dot{2}}=1\, ,\qquad \ep^{\a\b} \ep_{\b\gamma}=\delta^\a_\gamma  
\end{align}
and of a four-vector as a two-by-two matrix,
\begin{align}\label{}
x_{\a\dot{\alpha}}=x^\mu(\sigma_\mu)_{\a\dot{\alpha}}\,, \qquad \tilde x^{\da\a} = x^\mu(\tilde\sigma_\mu)^{\da\a} = \ep^{\a\b} \ep^{\da\db} x_{\b\db}\,.
\end{align}
These matrices satisfy the following identities (here $x \cdot y = x^\mu y_\mu$):
\begin{align}\label{}
& x_{\a\da} \tilde y^{\da\b} + y_{\a\da} \tx^{\da\b} = 2x \cdot y\, \delta_\a^\b\,, \qquad x_{\a\da} \tx^{\da\b} = x^2 \delta_\a^\b\,,\nt
&\tr(x \tilde y) = \tr(\tx y ) = 2 (x \cdot y)\,,\nt
& \tr(x \tilde y z \tilde t) = 2(x \cdot y) (z \cdot t) -  2(x \cdot z) (y \cdot t) + 2(x \cdot t) (y \cdot z) + 2i \ep(x,y,z,t) \,,\nt
& \tr(\tx y \tilde z t) = 2(x \cdot y) (z \cdot t) -  2(x \cdot z) (y \cdot t) + 2(x \cdot t) (y \cdot z) - 2i \ep(x,y,z,t)\,,
\end{align}
where $\ep(x,y,z,t) = \ep^{\mu\nu\la\rho} x_\mu y_\nu z_\la t_\rho$ and $\ep^{0123}=-1$.
We define the inverse matrix $x^{-1}$ by the relations
\begin{align}\label{}
(x^{-1})^{\da\a} = \frac{\tx^{\da\a}}{x^2}\,, \qquad x_{\a\da} (x^{-1})^{\da\b} = \delta^\b_\a \quad {\rm and} \quad (x^{-1})^{\da\a}x_{\a\db} = \delta^\da_\db\,.
\end{align}

\subsection{Conformal inversion and infinitesimal boosts}

It is well known that the conformal group $SO(2,4)$ can be generated by two operations, translation and inversion. In particular, a special conformal  transformation (boost) can be viewed as a succession of inversion, translation and another inversion, $K=IPI$. Under inversion a spacetime point transforms as follows, 
\begin{align}\label{}
I[x^\mu] = \frac{x^\mu}{x^2} \quad \Leftrightarrow \quad I[x_{\a\da}] = (x^{-1})^{\da\a}\,.
\end{align}
The difference of two points $x_{ij}=x_i-x_j$ is translation invariant and has the homogeneous inversion law (from here on we do not display the spinor indices) 
\begin{align}\label{}
I[x_{ij}] = -x^{-1}_i x_{ij} x^{-1}_j\,.
\end{align}
We deduce that strings of matrices with consecutive labels also transform covariantly, e.g, 
\begin{align}\label{1.4}
I[x_{ij}\tx_{jk}] = (x^2_j)^{-1}\, x^{-1}_i x_{ij} \tx_{jk} \tx^{-1}_k\,.
\end{align}
This is however not true for strings with a label gap, e.g., $x_{12}\tx_{35}$, etc. 

We can use the inversion law to obtain a finite conformal boost transformation.  After the first inversion we make a finite shift with parameter $B$ followed by another inversion:
\begin{align}\label{}
x_{ij}& \ \stackrel{I}{\longrightarrow} \ -x^{-1}_i x_{ij} x^{-1}_j \ \stackrel{P_{B}}{\longrightarrow} \ -(x_i+B)^{-1} x_{ij} (x_j+B)^{-1} \nt
& \stackrel{I}{\longrightarrow} \ -(x_i^{-1}+B)^{-1}x^{-1}_i x_{ij} x^{-1}_j (x_j^{-1}+B)^{-1}   =-(\mathbb{I}+x_i \tilde B)^{-1} x_{ij} (\tilde{\mathbb{I}}+\tilde B x_j )^{-1}\,.
\end{align}
Here $\mathbb{I} \equiv \delta_\a^\b$ and $\tilde{\mathbb{I}}  \equiv \delta_\da^\db$.
From this we derive the infinitesimal transformation with parameter $b\to0$:\begin{align}\label{14}
\delta_b x_{ij} =  x_i  \tb x_{ij} +x_{ij}  \tb x_j   \,.
\end{align} 
For covariant strings like in \p{1.4} we find
\begin{align}\label{15'}
\delta_b (x_{ij} \tx_{jk}) &= (x_i  \tilde b x_{ij} +x_{ij}  \tilde b x_j )\tx_{jk} + x_{ij} (\tx_j b \tx_{jk} + \tx_{jk} b \tx_k)\nt
& = x_i\tilde b \, x_{ij} \tx_{jk} + x_{ij} \tx_{jk}\, b \tx_k + (2b\cdot x_j) x_{ij} \tx_{jk} \,.
\end{align}
Observe that this homogeneous transformation involves matrix weights at the end points $i,k$ and a scalar weight at the middle point $j$. The trace of an even number of matrices is covariant if there are no label gaps between neighboring matrices, e.g.
\begin{align}\label{1.8}
& \frac1{2} \delta_b \tr(x_{ij} \tx_{ij}) = \delta_b (x^2_{ij}) =  2b\cdot(x_i+x_j) \, x^2_{ij}\,,\\
&\delta_b \tr(x_{ij} \tx_{jk} x_{kl} \tx_{li}) = 2b\cdot(x_i+x_j+x_k+x_l) \tr(x_{ij} \tx_{jk} x_{kl} \tx_{li})\,. \label{1.9}
\end{align}
From \p{1.8} we see that the notion of a lightlike vector, $x^2_{ij}=0$, is conformal.

Of course, the same results can be obtained with the four-vector form of the conformal boost generator \p{bK},
but we find the matrix composition rules more convenient and transparent. 

\subsubsection{Spinor-helicity variables}

Lightlike vectors, e.g. the on-shell momenta $p_i$, can be expressed in terms of spinor-helicity variables defined by the standard relation
\begin{align}\label{}
(p_i)_{\a\da} = (x_{i+1, i})_{\a\da} =\la_{i\, \a} \tl_{i\,\da} \equiv \ket{i}[i|\,.
\end{align}
Using  the transformation law \p{14}  we obtain 
\begin{align}\label{}
\delta x_{i+1\,, i} = x_{i+1} \tb \ket{i} [i| + \ket{i} [i|\tb x_i\,,
\end{align}
from where we can read off the transformations of the spinor-helicity variables
\begin{align}\label{a15}
\delta \ket{i} = x_{i+1} \tb \ket{i} \,, \qquad \delta [i| = [i|\tb x_i\,.
\end{align}

Then, other natural dual-conformal covariants are formed by strings of even or odd numbers of matrices, sandwiched between a pair of helicity spinors which compensate the transformations at the end points of the string, 
\begin{align}
\frac{\delta_b \, \bra{i} x_{ij_1} \tx_{j_1 j_2} \ldots \tx_{j_{n-1} j_n} x_{j_n k} |k]} { \bra{i} x_{ij_1} \tx_{j_1 j_2} \ldots \tx_{j_{n-1} j_n} x_{j_n k} |k]} = 2b \cdot ( x_{i+1} + x_{j_1} + \ldots + x_{j_n} + x_{k})
\,, \nt 
\frac{\delta_b \, \bra{i} x_{ij_1} \tx_{j_1 j_2} \ldots x_{j_{n-1} j_n} \tx_{j_n k} \ket{k}}{\bra{i} x_{ij_1} \tx_{j_1 j_2} \ldots x_{j_{n-1} j_n} \tx_{j_n k} \ket{k}} = 2b \cdot ( x_{i+1} + x_{j_1} + \ldots + x_{j_n} + x_{k+1})\,.
\end{align}

\subsection{Conformal boosts along a lightlike direction}

If strings of matrices with a label gap are not conformally covariant in general, in Ref.~\cite{Bern:2018oao} it was proposed to consider a subgroup of the conformal group, under which  such objects are still covariant. To illustrate the idea, consider the lightlike vector\footnote{We keep the convention of Ref.~\cite{Bern:2018oao} to use leg $p_3$ for creating nonplanar topologies, but we prefer a different labeling of the dual points, see \p{2.1}, in which $p_3=x_{43}$.} $x_{43}$,  with $x_{43}^2=0$, and make  the special choice of conformal boost parameter $b = \vep x_{43}$ with $\vep\to0$. Then from \p{14} we obtain  
\begin{align}\label{15}
\delta_{\vep x_{43}} x_{43} =\vep x_{43}^2(x_4+x_3) \ \stackrel{x_{43}^2=0}{\longrightarrow} \  0\,.
\end{align}
This result can be rewritten in terms of the conformal boost generator $K_\mu$ projected with the vector $x^\mu_{43}$:
\begin{align}\label{}
\left(x_{43}\cdot K\right) \, x^\mu_{43}=0 \quad {\rm if} \quad x_{43}^2=0 \,.
\end{align}
Due to this property the infinitesimal conformal transformations along the lightlike direction $x^\mu_{43}$ can be exponentiated, so they form a group. The generators of the conformal group, translation $P$, boost $K$, Lorentz  rotation $L$ and dilatation $D$ satisfy the algebra $[P_\mu, K_\nu] = L_{\mu\nu} + \eta_{\mu\nu} D$. The projection with $x^\mu_{43}$ defines the subalgebra $[P_\mu, \hat K_{43}] = \hat L_{\mu\, 43} + (x_{43})_\mu D$, where $\hat K_{43}=x_{43}\cdot K$, etc.

Ref. \cite{Bern:2018oao} proposes to study the conformal properties of nonplanar integrals obtained from planar ones by attaching a single additional leg ($p_3$ in their convention). Then they introduce dual coordinates that also include the new leg.  For example, for nonplanar configurations with five massless legs we can choose
\begin{align}\label{2.1}
& p_i = x_{i+1\, i} \quad {\rm with} \ x_6 \equiv x_1  \qquad  \Rightarrow \sum_{i=1}^5 p_i =0   \,.
\end{align}
The loop momenta are represented by internal dual points,  e.g., $x_6$ and $x_7$ in Fig.~\ref{fig:dualcoords}. 
In the planar case all the propagators can be put in the dual form $1/x_{ij}^2$, in the nonplanar case this is not possible. 
Then the main claim of Ref.~\cite{Bern:2018oao} is that for certain nonplanar integrals one can preserve part of the dual conformal symmetry. The typical situation occurs when the attachment of leg $p_3$ causes a shift of the momentum in some propagator by $p_3 = x_{43}$. After the shift (with a numerical factor $\gamma = \pm1$) the inverse propagator becomes
\begin{align}\label{113}
{x_{ij}^2} \ \to \ \hat x_{ij}^2 :=  {(x_{ij}+\gamma x_{43})^2} = {x_{ij}^2 +2\gamma x_{ij} \cdot x_{43}} \,.
\end{align}
We want to show that the new `hatted' interval $\hat x_{ij}^2$  transforms exactly as the original one. From \p{15} we know that we need to transform only $x_{ij}$ according to \p{14}:
\begin{align}\label{18}
\delta_{\vep x_{43}} (2x_{ij} \cdot x_{43}) &=   \delta_{\vep x_{43}}  \tr(x_{ij} \tx_{43}) =\vep \tr[(x_i \tx_{43} x_{ij} +x_{ij}\tx_{43} x_j)\tx_{43}] \nt
& = 2\vep x_{43}\cdot (x_i+x_j) \, (2x_{ij} \cdot x_{43})\,,
\end{align}
exactly as the transformation of $x_{ij}^2$, see \p{1.8}. Hence the shifted inverse propagator is indeed covariant with the same weights as the unshifted. This is the key observation which allows us to construct directional dual conformal invariants (DDCI). 

\subsubsection{DDC transformations of spinor-helicity variables}

The transformations \p{a15} of the spinor-helicity variables with the special parameter $b=\vep \ket{3}[3|$ become 
\begin{align}\label{}
\delta \ket{i} = \vep x_{i+1}|3] \vev{3i} \,, \qquad \delta [i| = \vep [i3] \bra{3} x_i\,.
\end{align}
With this we find that most of the Lorentz invariant brackets $\vev{ij}= \la^\a_i \la_{\a\, j}$ and $[ij]= \tl_{\da\, i} \tl^\da_j$ transform covariantly (here $\kappa_i = 2\vep p_3\cdot x_i $, with $\kappa_3=\kappa_4$):
\begin{align}\label{b17}
&\delta\vev{12}=\kappa_3 \vev{12} & \delta [12] = \kappa_1  [12]\nt
&\delta\vev{13}=\kappa_2 \vev{13}   & \delta [13] = \kappa_1  [12] \nt
&\delta\vev{15}=\kappa_2 \vev{15}   & \delta [15] = \kappa_5  [15] \nt
&\delta\vev{23}=\kappa_3 \vev{23}   & \delta [23] = \kappa_2  [23] \nt
&\delta\vev{24}=\kappa_5 \vev{24}   & \delta [24] = \kappa_2  [24] \nt
&\delta\vev{34}=\kappa_5 \vev{34}   & \delta [34] = \kappa_3  [34] \nt
&\delta\vev{35}=\kappa_1 \vev{35}   & \delta [35] = \kappa_5  [35] \nt
&\delta\vev{45}=\kappa_1 \vev{45}   & \delta [45] = \kappa_3  [45] 
\end{align}

The brackets $\vev{14}$ and $\vev{25}$ and their conjugates $[14]$, $[25]$ are not covariant. 

\subsection{Constructing DDCI expressions}

\subsubsection{Invariant ratios and cross-ratios}\label{appA41}

We have seen that the simple covariants like $x^2_{ij}$ can be deformed by a shift  along the direction of $x_{43}$  without changing their directional transformation properties. This allows us to immediately construct the following DDCI:
\begin{align}\label{115}
\hat u_{ij} = \frac{\hat x^2_{ij}} {x^2_{ij}} = 1+\frac{2\gamma x_{ij} \cdot x_{43}}{x^2_{ij}}= 1+\gamma \frac{\bra{3} x_{ij}|3]}{x^2_{ij}}\,, \qquad    \left(x_{43}\cdot K\right) \,  \hat u_{ij} = 0  \,,
\end{align}
for every $x^2_{ij}\neq0$.
The same logic applies to the trace of longer strings like 
\begin{align}\label{}
&\bra{3} x_{ij_1} \tx_{j_1 j_2} \ldots \tx_{j_{n-1} j_n} x_{j_n k} |3] = \tr (x_{ij_1} \tx_{j_1 j_2} \ldots \tx_{j_{n-1} j_n} x_{j_n k}\tx_{43} )\,.
\end{align}
They transform covariantly with weight $2x_{43}\cdot (x_i+x_k)$ at the end points, in addition to the usual weights $2x_{43}\cdot (x_{j_1}+\ldots +x_{j_n})$ at the internal points (see \p{15'}). Such strings can serve as numerators of DDCI integrals. There the conformal weights are balanced by a suitable denominator made of (possibly deformed as in \p{113}) propagators.  Other invariants, but this time carrying helicity, are obtained as products and ratios of the various brackets in \p{b17}.

The number of independent  DDCI variables can be determined as follows. For 5 lightlike momenta with momentum conservation one has 5 independent kinematical invariants $s_{ij} = 2p_i\cdot p_j$, or 4 dimensionless ratios of them. The single condition of DDCI \p{115} can eliminate one of them. So, it is always possible to find 3 independent variables that satisfy \p{115}. In practice, we can take any 3 of the $\hat u_{ij}$ as the independent DDCI.

\begin{figure}
\begin{center}
\includegraphics[width = 6cm]{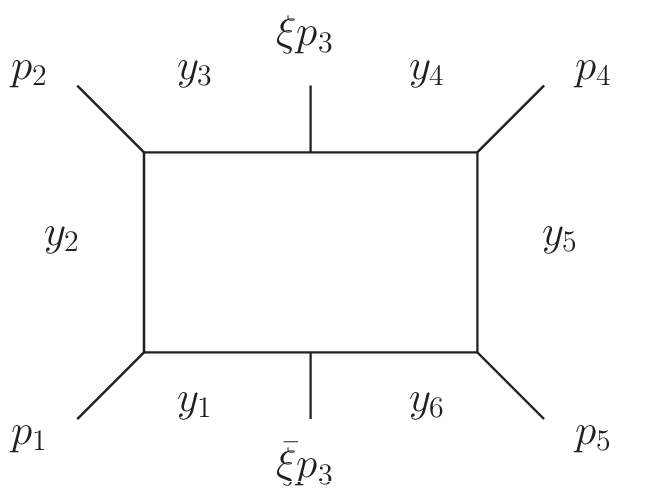}
\end{center}
\caption{ Six-point dual space for the hexagon integral \p{IaPole}. } \label{6pty}
\end{figure}

Let us compare this type of DDCI to the familiar cross-ratios made from four points,
\begin{align}\label{}
u_{ijkl} = \frac{x^2_{ij} x^2_{kl}}{x^2_{ik} x^2_{jl}}\,.
\end{align} 
Clearly, the condition for the existence of such invariants is that all the relevant $x^2_{pq}\neq0$. In particular, in a five-particle kinematics like \p{2.1} no such cross-ratios exist. Still, we can define DDCI  of the type \p{115}. We can also define DDCI cross-ratios
\begin{align}\label{ea27}
\hat u_{ijkl} = \frac{\hat x^2_{ij} \hat x^2_{kl}}{\hat x^2_{ik} \hat x^2_{jl}}\,.
\end{align}
This definition makes sense even if, e.g., $x^2_{ij}=0$ because the deformed interval $\hat x^2_{ij}\neq0$. Such invariants appear in our discussion of the hexagon integral \p{IaPole}. This planar integral has a dual space description with six dual points $y_1,\ldots,y_6$ shown in Fig.~\ref{6pty}. Being fully dual conformal, the integral is  a function of the 3 cross-ratios \p{3.14}. In terms of the dual coordinates $y_i$ they read
\begin{align}\label{a28}
u_1=\frac{y_{13}^2 y_{46}^2}{y_{14}^2 y_{36}^2}\,, \qquad 
u_2=\frac{y_{15}^2 y_{24}^2}{y_{14}^2 y_{25}^2}\,, \qquad 
u_3=\frac{y_{26}^2 y_{35}^2}{y_{25}^2 y_{36}^2}\,. 
\end{align} 
Now we write out the $y$'s in terms of the momenta and reexpress all the intervals $y^2_{pq}$ in terms of the 5-point dual coordinates $x_i$. Some intervals do not change, $y^2_{13}=x^2_{13}$, $y^2_{46}=x^2_{46}$. Others become deformed intervals (see \p{113}), e.g., $y^2_{14}=\hat x^2_{13}$. In particular, we get deformed intervals that would vanish if there was no deformation, e.g., $y^2_{15}=\hat x^2_{15}$. As a result, the 6-point cross-ratios \p{a28} become 5-point directional cross-ratios,
\begin{align}\label{a29}
u_1=\frac{x_{13}^2 x_{46}^2}{\hat x_{13}^2 \hat x_{46}^2}\,, \qquad 
u_2=\frac{\hat x_{15}^2 \hat x_{23}^2}{\hat x_{13}^2 \hat x_{25}^2}\,, \qquad 
u_3=\frac{\hat x_{12}^2 \hat x_{45}^2}{\hat x_{25}^2 \hat x_{14}^2}\,. 
\end{align} 
This explains why the integral \p{IaPole} is a DDCI. 

The leading pole residue \p{IbPole} involves some new elements whose DDCI properties require a comment. Firstly, $s_{12}=\vev{12}[21]$ and $s_{45}=\vev{45}[54]$ have  weights $\kappa_1+\kappa_3$  according to Table \p{b17}. Secondly, the quantities $t_{ijkl}$ are made from the `good' brackets from the table \p{b17} and have the same weights as the accompanying $s-$terms in the denominators in \p{IbPole} (see \p{18}), namely $-(\kappa_1+\kappa_2+\kappa_3+\kappa_5)$. Finally, the prefactor $\bra{2} p_4|5]=\vev{24}[45]$ has weight $\kappa_3+\kappa_5$. This results in the total weight $-(\kappa_1+\kappa_2)$ of the integral \p{IbPole}, which is also the weight of the result of the $\xi-$integration in \p{3.17}. We conclude that both forms of the residue $I^{(3)}_{a_2}$  are DDC covariant.

\subsubsection{Chiral invariants}\label{a42}

The planar pentagon alphabet has been introduced in \cite{Gehrmann:2015bfy} and extended to the 31-letter nonplanar case in \cite{Chicherin:2017dob},
\begin{align}\label{a30}
& W_{i} = s_{i,i+1} \;\;,\;\; W_{5+i} = s_{2+i,3+i} + s_{3+i,4+i} \;\;,\;\; W_{10+i} = s_{i,i+1} - s_{3+i,4+i} \,, \nt
& W_{15 + i} = - s_{i,i+2} \;\;,\;\; W_{20+i} = s_{i,i+2} + s_{i+2,i+3} \,, \nt
& W_{25+i} = \frac{\tr(\tilde p_{3+i} p_{4+i} \tilde p_{5+i} p_{6+i})}{\tr(p_{3+i} \tilde p_{4+i} p_{5+i} \tilde p_{6+i})} 
\;\;,\;\; W_{31} = 4 i \epsilon(p_1,p_2,p_3,p_4) \,, \quad \text{for} \;\; i=1,\ldots,5.
\end{align}
The letters here are split in groups of five related by cyclic permutations, except for the cyclic  invariant $W_{31}$. The two-particle invariants are $s_{jk} = 2 (p_j \cdot p_k)$ and the particle momenta are enumerated cyclically, $p_{6} \equiv p_1$.
The five letters $W_{26},\ldots,W_{30}$ are parity odd, and the remaining 26 letters are parity even. The letters $W_{1},\ldots,W_5$ and $W_{16},\ldots,W_{20}$ are admissible first entries of the symbols representing Feynman graphs.

It is natural to look for a set of 3 independent DDCI  among these letters. It turns out that the simplest choice are the chiral letters 
\begin{align}\label{21}
W_{26}   = \frac{[24]\vev{45}[51]\vev{12}}{\vev{24}[45]\vev{51}[12]}\,, \quad W_{27}  = \frac{[35]\vev{51}[12]\vev{23}}{\vev{35}[51]\vev{12}[23]}\,, \quad W_{30}     = \frac{[13]\vev{34}[45]\vev{51}}{\vev{13}[34]\vev{45}[51]}
\,.
\end{align}
The numerators and denominators differ by the chirality of the trace (see \p{a30}), i.e. they are complex conjugate, so these letters are pure phases. Therefore it is enough to show, with the help of the table \p{b17}, that the denominators are covariant, the numerators transform with the same (real) weights and the ratios are invariant.\footnote{We point out that  these letters satisfy additional DDCI condition, $(p_5\cdot K) W_{26} =0$,  $(p_5 \cdot K) W_{30} =0$ and $(p_1 \cdot K) W_{27} = 0$. This explains why only the letters $W_{26}$ and $W_{30}$ appear in the leading pole \p{330} of the integral $I_{(i)}$ with enhanced permutation symmetry.}

The set of pure phase letters    contains two more members, $W_{28}$ and $W_{29}$.
They involve the noncovariant brackets $\vev{14}$ and $\vev{25}$ and hence are not DDCI in the sense of $(p_3\cdot K)$, but are invariant under other projections of $K$.

\section{Finite six-dimensional integrals with the DDCI}
\label{sec:six-dim-examples}

\begin{figure}
\begin{center}
\begin{tabular}{ccc}
\includegraphics[width = 4cm]{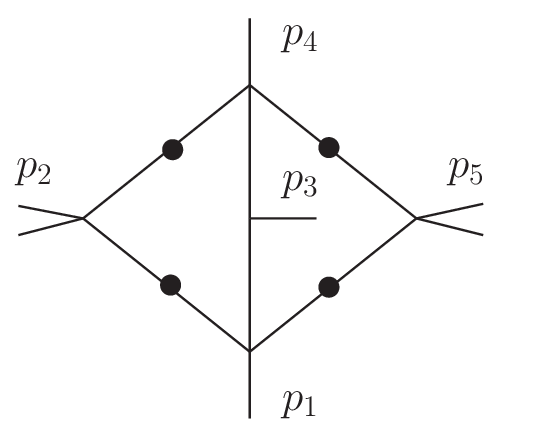} &
\includegraphics[width = 4cm]{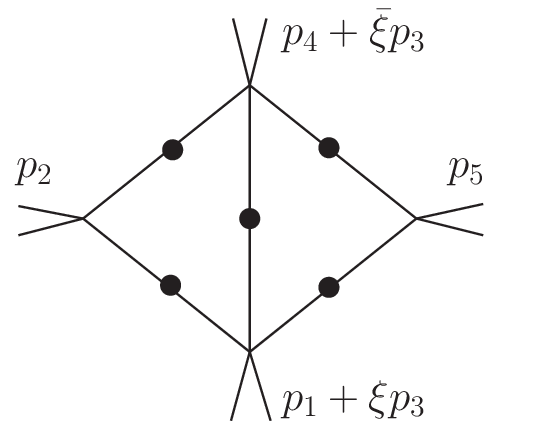} &
\includegraphics[width = 4.5cm]{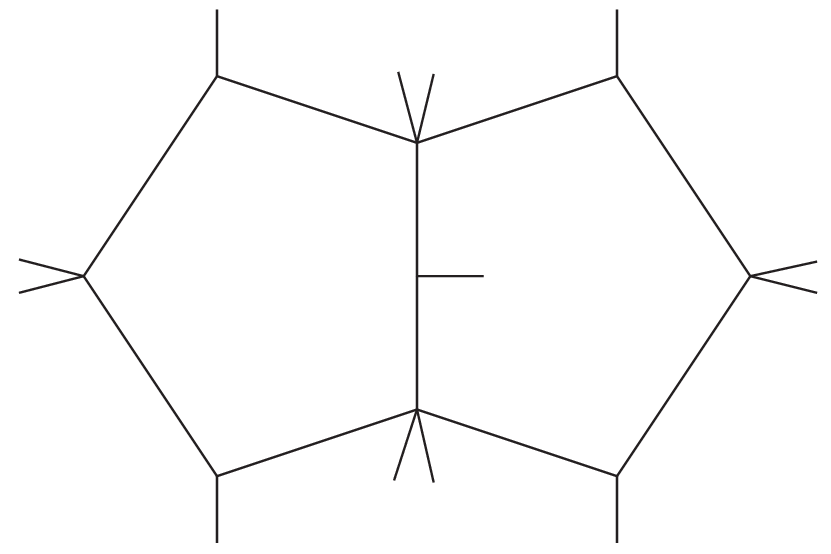} \\
(a) & (b) & (c)
\end{tabular}
\end{center}
\caption{ Finite six-dimensional integrals. Dots denote doubled propagators.} \label{finite6dintegrals}
\end{figure}

In this section we present examples of finite six-dimensional DDCI integrals.
We show a mechanism for finding an integral representation that makes the DDCI manifest.
This is achieved by writing the next-to-planar integral as a one-fold integral over a planar  DCI integral.
We also provide an example that has both the DDCI, as well as ordinary conformal symmetry.
This can be thought of as the analog of the `Yangian' invariant planar box integral \cite{Henn:2011xk}, and its generalizations, see e.g. \cite{Chicherin:2017frs}.

Consider the integral shown in Fig.~\ref{finite6dintegrals}(a). 
We call $I_{\ref{finite6dintegrals}(a)}$ the scalar integral, defined in six dimensions.
The kinematics is $p_{1}^2=p_{3}^2=p_{4}^2=0$, and $p_2^2 \neq 0, p_5^2 \neq 0$.
The integral is finite, both in the ultraviolet and in the infrared. 
It is next-to-planar, with the light-like leg $p_{3}$ leading to the nonplanarity. 
By power counting it is simple to see that each subintegral is dual conformal. Hence it is DDCI under $(p_{3} \cdot K)$. 

For integrals of this type, that do not involve any loop-dependent numerator factors, it is straightforward to make the DDCI manifest.
This can be seen by relating the integral to a planar integral.
We achieve this via the standard trick of Feynman-combining two propagators adjacent to the on-shell leg $p_3$ (see also \p{gelf}),
\begin{align}
\frac{1}{\ell^2 (\ell+p_{3})^2}  = \int_0^1 d\xi \frac{1}{[ (\ell+ \xi p_{3})^2]^2} \,.
\end{align}
In this way, we obtain the integral representation
\begin{align}\label{integralrep1}
I_{\ref{finite6dintegrals}(a)} = \int_0^1 d\xi\, I_{\ref{finite6dintegrals}(b)}(u,v) \,.
\end{align}
This formula relates the next-to-planar integral $I_{\ref{finite6dintegrals}(a)}$ to the planar integral shown in Fig.~{\ref{finite6dintegrals}(b)}. The latter is dual conformal,
\begin{align}
I_{\ref{finite6dintegrals}(b)} = \frac{1}{(y_{14}^2 y_{23}^2 )^2} \tilde{I}_{\ref{finite6dintegrals}(b)}(u,v)  \,,
\end{align}
and thus depends on two dual conformal cross-ratios $u,v$.
In Eq.~(\ref{integralrep1}) they are parametrized by $\xi$ in the following way,
\begin{align}\label{xideformation}
u = \frac{y_{12}^2 y_{34}^2}{y_{14}^2 y_{23}^2} \,,\quad v =  \frac{y_{13}^2 y_{24}^2}{y_{14}^2 y_{23}^2}  \,, 
\end{align}
with
\begin{align}
y_{12}^2 =& (p_1 + p_3 \xi)^2  \,, \nonumber \\ 
y_{34}^2 =& (p_4 + p_3 \bar{\xi})^2  \,, \nonumber \\ 
y_{13}^2 =& (p_4+p_5 + p_3 \bar{\xi})^2  \,, \nonumber \\ 
y_{24}^2 =& (p_5 +p_1 + p_3 {\xi})^2  \,, \nonumber \\ 
y_{14}^2 =& p_5^2 \,, \nonumber \\ 
y_{23}^2 =& p_{2}^2 \,.
\end{align}
The integral $\tilde{I}_{\ref{finite6dintegrals}(b)}$ can be expressed in terms of a one-loop integral via the conformal star-triangle relation. Finally, the resulting one-loop integral can be calculated by standard methods. 
It is convenient to express the answer in the following variables
\begin{align}
u = z_1 z_2 \,,\qquad v = (1-z_1 ) (1-z_2) \,.
\end{align}
We find
\begin{align}
\tilde{I}_{\ref{finite6dintegrals}(b)}(u,v) = h(z_1,z_2) = \sum_{i=1}^{3} r_i h_i\,,
\end{align}
where 
\begin{align}
r_1 = \frac{z_1 + z_2 - 2 z_1 z_2}{(z_1 - z_2)^3}
\,,\quad
r_2  = \frac{z_1 + z_2}{(z_1 - z_2)^2}
\,,\quad
r_3 = \frac{2 - z_1 - z_2}{(z_1 - z_2)^2} 
\end{align}
and
\begin{align}
h_1 =&   2 {\rm Li}_{2}(z_1 ) -  2 {\rm Li}_{2}(z_2 ) +\left[ \log(z_1 z_2) \log(1-z_1)-\log(1-z_2) \right]\,,  \nonumber\\
h_2 =&  \log( z_1 z_2 ) \,, \nonumber\\
h_3 =&  \log( (1-z_1) (1-z_2) )\,.
\end{align}
As a check, we mention in passing that the integral satisfies  the D'Alembert equation \cite{Drummond:2006rz},
which in the present case takes the form
\begin{align}
2
   h^{(0,1)}(z_{1},z_{2})-2 h^{(1,0)}(z_{1},z_{2})+(z_{1}-z_{2})
   h^{(1,1)}(z_{1},z_{2})=- \frac{z_1 - z_2}{ z_{1} (1-z_{1}) z_{2} (1-z_{2}) } \,. 
\end{align}
We comment that when using Eq.~(\ref{integralrep1}) in practice, care has to taken to keep track of the $i0$ prescription of the Feynman propagators.

From the discussion in Appendix \ref{aA} we know that the deformation along $p_3$ of the dual conformal cross-ratios in Eq.~(\ref{xideformation}) preserves the DDCI. 
We conclude that for next-to-planar integrals without numerator factors, the DDCI follows straightforwardly from the ordinary dual conformal symmetry of an associated planar integral.

Finally, we mention a particularly interesting class of finite integrals that have both DDC invariance, as well as ordinary conformal symmetry \cite{Chicherin:2017bxc}. There are many such integrals. Here, we give one example that may be relevant for seven- and higher-particle scattering amplitudes, see Fig.~\ref{finite6dintegrals}(c).
Given the structure of the massless corners, we may relate this integral to the planar integral that we just computed. In complete analogy with the above discussion, we can introduce five $\xi$ variables, yielding the integral representation
\begin{align}\label{integralrep2}
I_{\ref{finite6dintegrals}(c)} = \int_0^1 \left( \prod_{i=1}^{5} d\xi_{i} \right) \, I_{\ref{finite6dintegrals}(b)}(u,v) \,,
\end{align}
with the parametrization of $u,v$ following from the kinematics of Fig.~{\ref{finite6dintegrals}(c)}.

\providecommand{\href}[2]{#2}\begingroup\raggedright\endgroup

\end{document}